\documentclass{webofc}

\usepackage[varg]{txfonts}   
\usepackage{hyperref}
\usepackage{url}
\hypersetup{colorlinks=true,citecolor=blue,urlcolor=blue,linkcolor=blue}
\usepackage{amsmath}
\usepackage{amssymb}
\usepackage{physics}
\usepackage[left]{lineno}
\begin{document}
%
\title{Effects of hadronic reinteraction on jet fragmentation from small to large systems}
%
%

\author{\firstname{Hendrik} \lastname{Roch}\inst{1}\fnsep\thanks{\email{Hendrik.Roch@wayne.edu}}
        \firstname{} \lastname{for the JETSCAPE collaboration}        
}

\institute{Department of Physics and Astronomy, Wayne State University, Detroit MI 48201}

\abstract{
We investigate the impact of the hadronic phase on jet quenching in nuclear collider experiments, an open question in heavy-ion physics. 
Previous studies in a simplified setup suggest that hadronic interactions could have significant effects, but a systematic analysis is needed. 
Using the X-SCAPE event generator with the SMASH afterburner, we study the role of hadronic rescattering on jet fragmentation hadrons. 
Applying this framework to $e^++e^-$ collisions, we demonstrate that even in small systems with limited particle production, hadronic interactions lead to measurable modifications in final-state hadronic and jet observables by comparing scenarios with and without afterburner rescattering.
}
\maketitle
\section{Introduction}
\label{intro}
The JETSCAPE event generator provides a modular and task-based framework for simulating various aspects of heavy-ion collisions~\cite{Putschke:2019yrg}.
For this study, we use a novel hadronization module, Hybrid Hadronization~\cite{Han:2016uhh,Fries:2019vws}, which implements a natural interpolation between quark recombination~\cite{Fries:2003vb,Fries:2003kq} in dense phase-space regions and Lund string fragmentation~\cite{Andersson:1983ia} in dilute regions.
The Hybrid Hadronization module provides the full phase-space information of the produced hadrons, allowing us to input these hadrons into the hadronic afterburner SMASH~\cite{SMASH:2016zqf} for further propagation until kinetic freeze-out is reached.
An exploratory study in a simplified setup of a radially expanding hadron gas with a beam of pions has shown measurable effects of hadronic reinteractions on high-momentum hadrons, including the transfer of energy to larger angles away from the jet axis~\cite{Dorau:2019ozd}.
With the implementation of this physics in the X-SCAPE framework, we can study these effects in more detail and quantify their impact on final-state hadronic and jet observables.

\section{Simulation setup}
\label{sec:setup}
In this study, we use the upcoming X-SCAPE 1.2 with an $e^++e^-$ gun module involving PYTHIA 8.3~\cite{Bierlich:2022pfr} to generate the hard scattering at $\sqrt{s}=91.2\;\mathrm{GeV}$.
The produced hard parton system is then passed to MATTER for subsequent showering until the parton virtuality drops below a cutoff scale $Q_0$~\cite{Majumder:2013re,Cao:2017qpx}.
Partons below $Q_0$ are then hadronized with Hybrid Hadronization, and the resulting hadrons are fed into SMASH for further propagation.

\subsection{Hybrid Hadronization}
\label{subsec:hybrid_hadronization}
The combination of quark recombination and string fragmentation allows Hybrid Hadronization to hadronize parton showers from vacuum systems like $e^++e^-$ and $p+p$ as well as from systems that include a hydrodynamic medium in $A+A$ collisions.
Both quark recombination and string fragmentation of shower partons in Hybrid Hadronization can utilize thermal partons from the background medium.
In the following, we outline the steps performed in the Hybrid Hadronization module.

The input for Hybrid Hadronization is a list of partons from energy-loss modules where the virtuality is below $Q_0$.
Prior to the recombination step, all gluons are provisionally forced to decay into quark-antiquark pairs non-perturbatively.
This step is not finalized if neither the quark nor the antiquark daughter from the decay is used in the subsequent recombination stage.
In such cases, the gluon is returned to the parton list at the end of the recombination step.
Additionally, if a hydrodynamic medium is present, the freeze-out surface is used to sample thermal up, down, and strange quarks, following a Cooper-Frye sampling prescription.
These thermal partons serve as a reservoir with which the shower partons can interact, randomizing the color flow among them since partons,~e.g., from LBT, have color tags set to zero, indicating random color.
The module then loops through all shower partons and checks for recombination with either another shower parton or a thermal parton if a hydrodynamic medium is present.

The recombination routine is based on a Wigner function formalism, where the recombination probability is determined by the overlap of Wigner functions of the partons.
Wigner functions of mesons and baryons are computed assuming a 3D isotropic harmonic oscillator potential between quarks, with potential sizes derived from data on squared charge radii.
To compute the phase space probability $P_{\rm ps}$ to form a meson/baryon, its Wigner distribution $W_{M}(\mathbf{x},\mathbf{p})$/$W_{B}(\mathbf{x}_1,\mathbf{p}_1;\mathbf{x}_2,\mathbf{p}_2)$ is 
convoluted with Gaussian wave packets representing the quark ($W_{q}(\mathbf{x},\mathbf{p})$)/antiquark ($W_{\bar{q}}(\mathbf{x},\mathbf{p})$) wave functions~\citep{Han:2016uhh,Kordell:2021prk}:
\begin{align}
P_{{\rm ps}, M}(\mathbf{r},\mathbf{q})&\sim\int\prod\limits_{i=q,\bar{q}}\mathrm{d}^3 x_i\mathrm{d}^3 p_i\; W_{M}(\mathbf{x},\mathbf{p}) W_q(\mathbf{x}_1,\mathbf{p}_1)W_{\bar{q}}(\mathbf{x}_2,\mathbf{p}_2),\\
P_{{\rm ps}, B/\bar{B}}(\mathbf{r},\mathbf{q})&\sim\int\prod\limits_{i=q/\bar{q}}\mathrm{d}^3 x_i\mathrm{d}^3 p_i\; W_{B/\bar{B}}(\mathbf{x}_1,\mathbf{p}_1;\mathbf{x}_2,\mathbf{p}_2) W_{q/\bar{q}}(\mathbf{x}_1,\mathbf{p}_1) W_{q/\bar{q}}(\mathbf{x}_2,\mathbf{p}_2) W_{q/\bar{q}}(\mathbf{x}_3,\mathbf{p}_3).
\end{align}
Here, $\mathbf{x}=\mathbf{x}_1-\mathbf{x}_2$ and $\mathbf{p}=(m_2\mathbf{p}_1-m_1\mathbf{p}_2)/(m_1+m_2)$ are the relative position and weighted momentum of the quarks forming the meson, while in the baryon case we define $\mathbf{y}_1$ and $\mathbf{k}_1$ similarly.
The other relative coordinate is then given by $\mathbf{x}_2=(m_1\mathbf{x}_1+m_2\mathbf{x_2})/(m_1+m_2)-\mathbf{x}_3$ and $\mathbf{p}_2=(m_3(\mathbf{p}_1+\mathbf{p}_2)-(m_1+m_2)\mathbf{p}_3)/(m_1+m_2+m_3)$ with $m_k$ denoting the constituent masses of the (anti)quarks.
The probabilities $P_{\rm ps}(\mathbf{r},\mathbf{p})$ depend on the relative phase space coordinates of the centroids of the (anti)quark wavepackets and are multiplied by the corresponding probabilities $P_{\rm spin}$ and $P_{\rm color}$ for overlap of the spin and color states, where the spin is always treated statistically.
If available, the color tag information of the partons is used in the recombination process.
The total probability to form a hadron is given by $P_{\rm tot}=P_{\rm ps}P_{\rm spin}P_{\rm color}$ in the center of mass frame of the possibly produced hadron.
If the hadron is sampled or not is then determined by sampling a random number and comparing this to $P_{\rm tot}$.
In case the hadron is formed, the partons are removed from the parton list, and the algorithm continues until either all partons are used or all possible combinations are rejected.

A key feature of Hybrid Hadronization for heavy-ion collisions is the ability to recombine shower partons with the thermally produced medium partons, allowing collective flow to be transferred to semi-hard hadrons.
The ability to smoothly switch on recombination processes if more dense QCD medium is present has been shown in Ref.~\cite{JETSCAPE:2025wjn}.

Partons that do not participate in recombination are treated as remnants in the Hybrid Hadronization module and are connected by strings using known color tags.
While the recombination step only removes color singlet states from the partonic system, the provided string system might not be complete at the beginning of the hadronization step, such that these incomplete systems are repaired with partons from the background medium if available or additional partons assumed close to the beam rapidity.
Each recombination baryon (antibaryon) creates an antijunction (junction) in order to conserve the net baryon number.
The remaining color singlet string system is then hadronized via Lund string fragmentation in PYTHIA.
The framework allows for either fully decaying excited hadrons within PYTHIA or transferring them to the hadronic afterburner SMASH for further rescattering.

\subsection{SMASH}
\label{subsec:SMASH}
SMASH (Simulating Many Accelerated Strongly-interacting Hadrons)~\cite{SMASH:2016zqf} is a hadronic transport approach that effectively solves the Boltzmann equation to simulate the non-equilibrium dynamics of hadrons in full phase space:
\begin{align}
p^\mu\partial_{\mu}f_i(\mathsf{x},\mathsf{p}) = \mathcal{C}^i_{\rm coll}.
\end{align}
Here, $\mathcal{C}^i_{\rm coll}$ is the collision term, and $f_i(\mathsf{x},\mathsf{p})$ is the single-particle phase space distribution for particle species $i$.
All hadronic states up to $m\leq 2.35\;\mathrm{GeV}$, along with their corresponding decays and cross sections, are implemented under the assumption of isospin symmetry.
In afterburner mode, SMASH is initialized with an external hadron list provided by X-SCAPE, containing hadrons produced by Hybrid Hadronization and, if hydrodynamics is included, also from soft particlization.

\section{Results}
In this article, we present results exclusively from $e^++e^-$ collisions, as they provide a clean environment for studying hadronic rescattering effects in the afterburner phase without the complication of an underlying event.
To investigate these effects, we conduct three types of simulations.

The first simulation serves as a baseline, where hadrons produced in Hybrid Hadronization are passed to SMASH, but only decays are performed.
The other two simulations introduce a short free-streaming period after hadronization before allowing rescattering in SMASH. 
We consider two free-streaming times, 0.1 fm and 1.0 fm, to account for the fact that hadronization may not occur instantaneously.
This variation, which is a free parameter of the model, affects the initial hadron density in the afterburner phase.

For each setup, we simulated a total of $1.5\times 10^6$ events using a preliminary parameter set from an ongoing Bayesian study. 
The analysis was carried out with the Python package \texttt{SPARKX}~\cite{hendrik_roch_2025_14931783,Sass:2025opk}. 
Events containing heavy-flavor hadrons were excluded, as SMASH does not currently implement their decay channels.

The first observable under study is the event shape variable thrust, $T$, defined as: $T\equiv\max_{\mathsf{n}_T}\left[\left(\sum_i\abs{\vec{p}_i\cdot\mathsf{n}_T}\right)/\left(\sum_i \abs{\vec{p}_i}\right)\right]$, where the sum runs over all selected particles in an event, and $\mathsf{n}_T$ is the unit vector that maximizes $T$.
We define ``thrust'' as $\tau \equiv 1 - T$, such that $\tau$ vanishes for pencil-like events.
The second event shape observable is the total jet broadening, $B_T$, given by: $B_\mathrm{T}=B_1+B_2$ where each hemisphere broadening is defined as: $B_k\equiv \left(\sum_{i\in H_k}\abs{\vec{p}_i\times\mathsf{n}_T}\right)/\left(2\sum_i\abs{\vec{p}_i}\right)$ and the sum runs over the particles in each hemisphere defined by the thrust axis.

\begin{figure*}
\centering
\includegraphics[width=0.49\textwidth,clip]{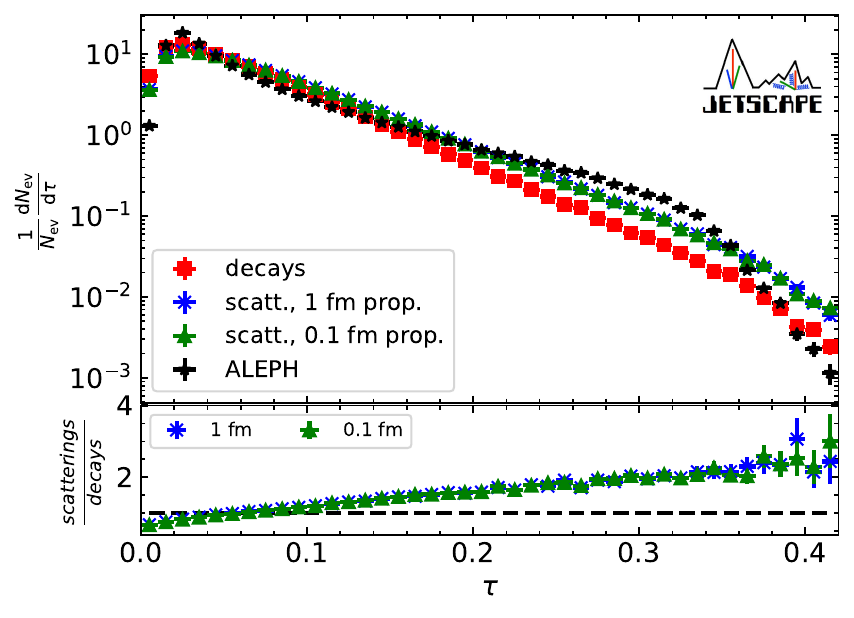}
\includegraphics[width=0.49\textwidth,clip]{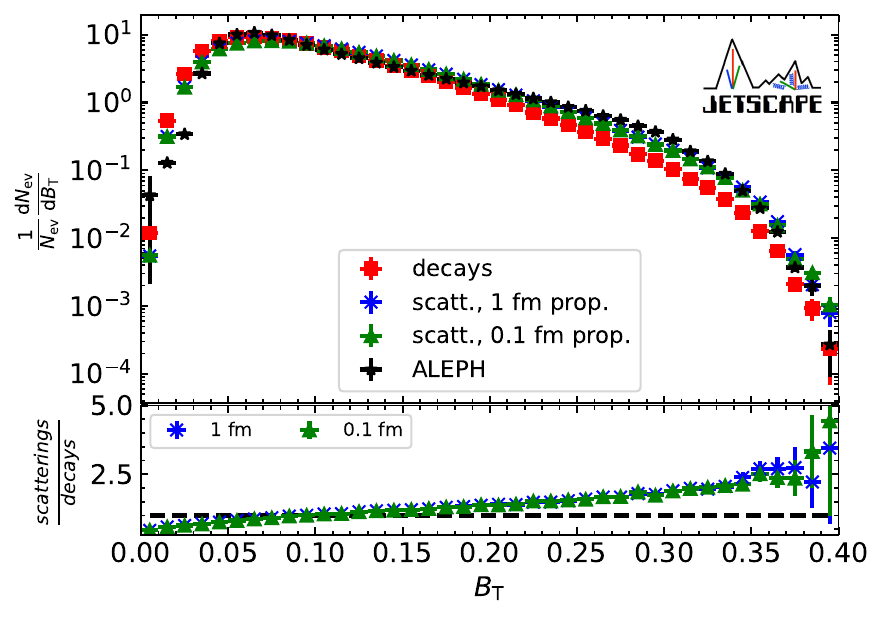}
\caption{Thrust $\tau$ (left) and total jet broadening $B_T$ (right). Squares represent the run with only decays, while crosses and triangles correspond to 1.0 fm and 0.1 fm free-streaming times, respectively. ALEPH data~\cite{ALEPH:2003obs} is shown for comparison (stars). The ratio plot compares results to the baseline decay-only run.}
\label{fig-1}
\end{figure*}

Figure~\ref{fig-1} (left) compares all three simulation setups and data from the ALEPH experiment~\cite{ALEPH:2003obs}.
We find that rescattering in the hadronic phase broadens events by scattering hadrons away from the thrust axis. 
The right panel, showing total jet broadening, confirms this effect, as events become wider in momentum space when rescattering is included. 
Notably, varying the free-streaming time before rescattering does not significantly impact the results. 
Additionally, the thrust distribution does not fully match experimental data at large $\tau$, which is not unexpected, as the event generator settings are not tuned using these observables. 
However, including hadronic final-state interactions improves the description of total jet broadening data.

\begin{figure*}
\centering
\includegraphics[width=0.49\textwidth,clip]{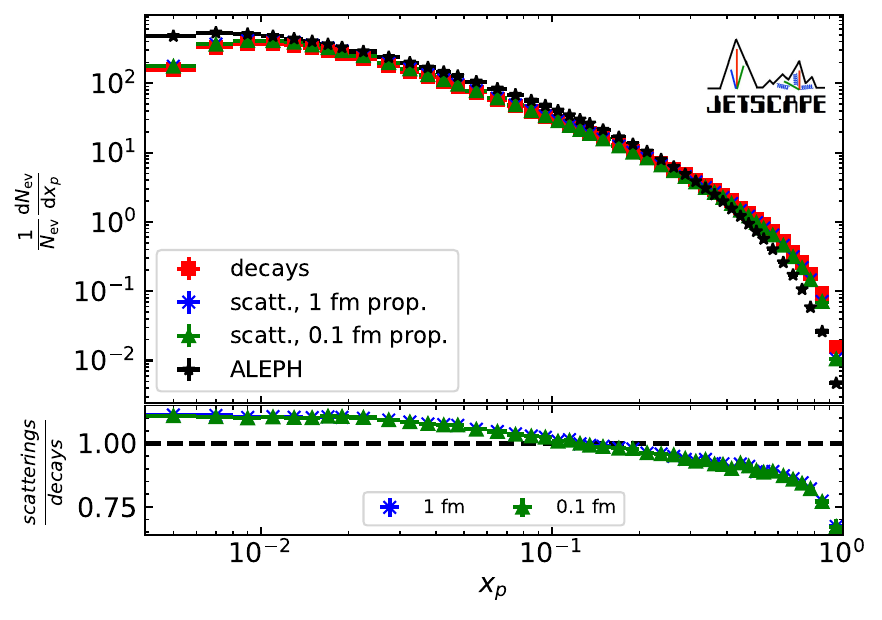}
\includegraphics[width=0.49\textwidth,clip]{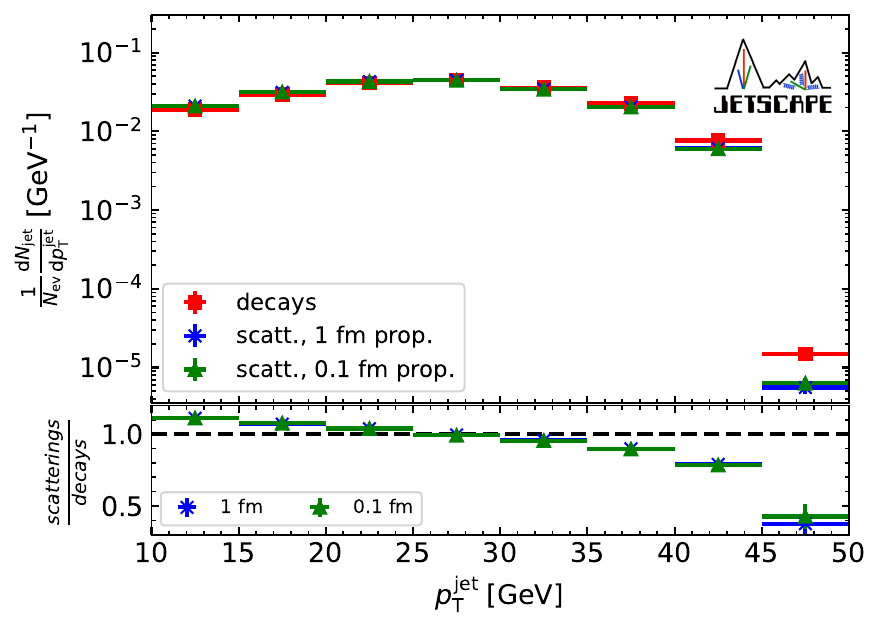}
\caption{Hadronic $x_p$ spectrum for charged hadrons compared to ALEPH data~\cite{ALEPH:1996oqp} (left) and transverse momentum spectrum of jets (right). The same symbols as in Fig.~\ref{fig-1} are used. The ratio plot compares results to the baseline decay-only run.}
\label{fig-2}
\end{figure*}

Figure~\ref{fig-2} (left) presents the hadronic $x_p\equiv 2|\vec{p}|/\sqrt{s}$ spectrum, where we observe that large-momentum hadrons lose energy due to scatterings, resulting in a diffusion of hadrons from high to low momenta.
A similar trend appears in the hadronic jet transverse momentum spectrum (Fig.\ref{fig-2}, right), where scatterings lead to a shift of high-momentum jets toward lower $p_T$. 
Jets were identified using the \texttt{FastJet}~\cite{Cacciari:2011ma} wrapper with an $e^++e^-$ jet algorithm, requiring $p_T \geq 10,\mathrm{GeV}$, $|\eta| \leq 1.74$, and a jet radius of $R = 0.8$. 
Again, the free-streaming time does not appear to affect the results significantly.

\begin{figure*}
\centering
\includegraphics[width=0.49\textwidth,clip]{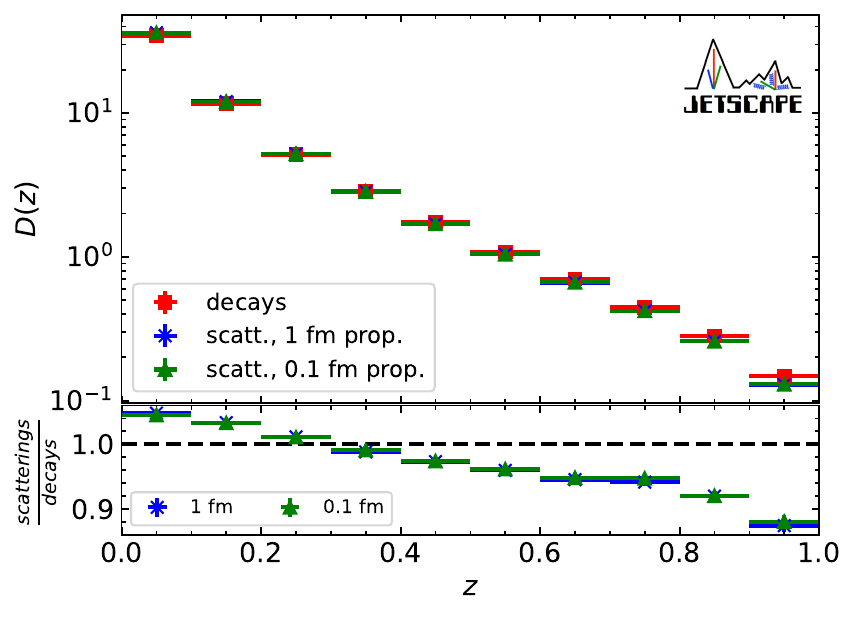}
\includegraphics[width=0.49\textwidth,clip]{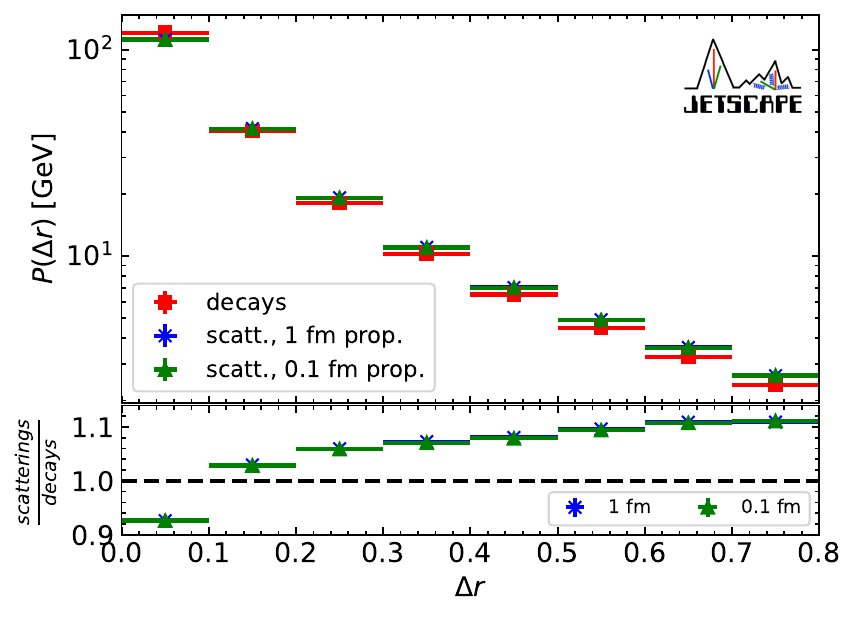}
\caption{Jet fragmentation function (left) and jet shape (right). The same symbols as in Fig.~\ref{fig-1} are used. The ratio plot compares results to the baseline decay-only run.}
\label{fig-3}
\end{figure*}

This momentum redistribution is further evident in the jet fragmentation function $D(z)$ (Fig.\ref{fig-3}, left), defined as $z \equiv p_T^{\mathrm{ch}} / p_T^{\mathrm{jet}}$, where high-momentum hadrons in the jet transfer momentum to lower-momentum particles.
The right panel of Fig.\ref{fig-3} shows the jet shape $P(\Delta r)$, with $\Delta r \equiv \sqrt{(\eta - \eta_{\mathrm{jet}})^2 + (\phi - \phi_{\mathrm{jet}})^2}$, demonstrating that rescatterings shift hadrons from the jet core to larger radii.

\section{Conclusion}
We have demonstrated that incorporating hard hadrons from Hybrid Hadronization into the hadronic afterburner SMASH significantly affects event shapes, hadronic spectra, and jet observables. 
Rescattering transfers momentum from high-energy hadrons to softer ones, broadens event shapes, and scatters hadrons away from the thrust or jet axis. 
These findings indicate sizable hadronic rescattering effects on hard hadrons even in the most dilute collider system, $e^++e^-$.

This study provides a foundation for future investigations of other systems, including $p+p$ and heavy-ion ($A+A$) collisions, where underlying event effects are present.
Furthermore, we aim to apply this framework to higher-energy $e^++e^-$ collisions, where significant two-particle correlations have been observed in high-multiplicity events that current event generators fail to describe~\cite{Chen:2023nsi}. 
The momentum reshuffling observed in the afterburner phase may help bridge the gap between theory and experiment.

\section{Acknowledgments}
These proceedings are supported in part by the National Science Foundation (NSF) within the framework of the JETSCAPE collaboration, under grant numbers ACI-1550300 OAC-2004571 (CSSI:X-SCAPE) and in part by the U.S. Department of Energy (DOE) under grant number DE-SC0024232.
%
\bibliography{bibliography} 

\end{document}